\documentclass[12pt]{article}
\usepackage{jheppubmod,amsmath,amssymb}
\pdfoutput=1
\usepackage{color}
\usepackage{amsmath,amssymb}
\usepackage{comment}
\usepackage{braket}
\usepackage{mathtools}
\usepackage{psfrag}
\usepackage{array}
\usepackage{amssymb}
\usepackage{amsmath}
\usepackage{amsthm}
\usepackage{graphicx}
\usepackage{epstopdf}
	
\usepackage{color}
\usepackage{epsfig}
\usepackage[punctsep]{collref}

\def\pb[#1,#2]{\{#1, #2\}}
\def\deb[#1,#2]{[#1,#2]_{\text{D.B.}}}

\def\tr{{\rm Tr}}

\def\Or[#1]{{\text{O}}\left({#1}\right)}
\def\dotl[#1,#2]{\left\langle #1,\, #2 \right\rangle}
\def\dotlb[#1,#2]{\left\langle #1,\, #2 \right\rangle}
\def\dotlm[#1,#2]{\left[ #1,\, #2 \right]}
\def\dotp[#1,#2]{(\vect{#1} \cdot\vect{#2})}
\def\aff[#1,#2]{\hat{#1}(#2)}

\def\n4sym{{\cal N}=4 SYM}
\def\>{\rangle}
\def\<{\langle}
\def\weight[#1,#2,#3]{\{(#1),#2,#3\}}
\def\ads[#1]{$\text{AdS}_{#1}$}

\def\tarelr[#1]{\widetilde{a}^{\text{rel}}_{R#1}}
\def\Oright[#1]{{\cal O}_{R#1}}
\def\Oleft[#1]{{\cal O}_{L#1}}
\def\aleft[#1]{a_{L#1}}
\def\arelr[#1]{a^{\text{rel}}_{R#1}}

\newcommand{\be}{\begin{equation}}
\newcommand{\ee}{\end{equation}}
\newcommand{\bea}{\begin{eqnarray}}
\newcommand{\eea}{\end{eqnarray}}
\newcommand{\ba}{\begin{align}}
\newcommand{\ea}{\end{align}}
\newcommand{\bs}{\begin{split}}
\def\sess\end{split}

\newcommand{\vect}[1]{{#1}}

\title{Connecting Fisher information to bulk entanglement in holography}

\author[a,b]{Souvik Banerjee,}
\emailAdd{souvik.banerjee@uu.se}

\affiliation[a]{Department of Physics and Astronomy, Uppsala University, \\
 SE-751 08 Uppsala, Sweden.\\ }

\author[b,c]{Johanna Erdmenger,}
\emailAdd{jke@mpp.mpg.de}
\affiliation[b]{Institut f\"ur Theoretische Physik und Astrophysik,
  Julius-Maximilians-Universit\"at W\"urzburg, \\   Am Hubland, 97074
  W\"urzburg, Germany.\\}
\affiliation[c] {Max-Planck-Institut f\"ur Physik (Werner-Heisenberg-Institut),\\
 F\"ohringer Ring 6, 80805 Munich, Germany.\\}
\author[c,d]{Debajyoti Sarkar}
\emailAdd{debajyoti.sarkar@physik.uni-muenchen.de}
\affiliation[d]{Arnold Sommerfeld Center,\\ 
Ludwig-Maximilians-University, Theresienstr. 37, 80333 Munchen, Germany.\\}

\vspace{3mm}


\abstract{In the context of relating AdS/CFT to quantum information theory, we
propose a holographic dual of Fisher information metric for mixed
states in the boundary field theory. This amounts to a holographic
measure for the distance between two mixed quantum states. For a spherical subregion in the boundary we show that this is related to a particularly regularized volume enclosed by the Ryu-Takayanagi surface.
We further argue that the quantum correction to the proposed Fisher information metric is related to the quantum correction to the boundary entanglement entropy.  We discuss consequences of this connection.}

\begin{document}
\preprint{UUITP-03/17, LMU-ASC 01/17}
\maketitle

\section{Introduction}\label{sec:backgnd}
For the last two decades, a significant amount of efforts have been devoted to realizing connections of quantum information theory \cite{Nielsenbook} to geometry and gravity. Within string theory, the first realization of such a connection is the calculation of black hole entropy through the counting of black hole microstates for supersymmetric black holes \cite{Strominger:1996sh}.  

These connections were put on an even stronger footing with the advent of AdS/ CFT correspondence \cite{Maldacena:1997re}. One of the most important and crucial steps forward in this direction is the holographic realization of entanglement entropy \cite{Ryu:2006bv, Hubeny:2007xt}.
The entanglement entropy for an entangling region in a conformal field theory living on the boundary of an asymptotically AdS spacetime is proposed to be given by a minimal codimension-two area in the bulk. The associated bulk hypersurface, also referred to as Ryu-Takayanagi (RT) surface, is homologous to the boundary entangling region.  In two dimensions, this holographic computation of boundary entanglement entropy matches exactly with the direct replica trick computation in  conformal field theory  \cite{Calabrese:2009qy}. Subsequently, a direct holographic justification of the RT formula was provided in \cite{Lewkowycz:2013nqa}. Moreover, the holographic realization of quantum entanglement is  extremely useful in understanding the structure of the dual bulk spacetime. For eternal black holes \cite{Maldacena:2001kr}, quantum entanglement was found to have a direct bulk interpretation in terms of regularity of the horizon \cite{VanRaamsdonk:2009ar} which turns out to be an extremely useful input in understanding the black hole information paradox.

A further concept from quantum information theory recently studied in the context of holography is the \emph{distance between two quantum states}. There exist two well-accepted ways to define the  distance between two generic quantum states, namely (i) the Fisher information metric and (ii) the Bures metric or fidelity susceptibility \cite{Nielsenbook, Uhlmann:1975kt}. 

In order to define these, let us consider a generic density matrix $\sigma$ and perturb it by some parameter $\lambda$ which parametrizes the quantum state. Here for simplicity we assume a one-parameter family of states which however can be generalized for arbitrary number of parameters. 
If $\rho$ is the new density matrix corresponding to the fluctuation $\lambda \rightarrow \lambda+ \delta \lambda$, then for nearby states, the density matrix $\rho$ can be expanded as
\be
\rho = \sigma + (\delta\lambda) \rho_1 +  \frac{1}{2} (\delta\lambda)^2\rho_2 + \cdots,
\ee
where for simplicity we have chosen the parametrization in a way such that the initial state $\sigma$ coincides with $\lambda =0$.
The coefficients $\rho_1$ and $\rho_2$ are of first and second order in $\delta \rho$, respectively, with $\delta \rho$ a small deviation from $\sigma$. In this case, a  distance metric may be defined by
\be
\label{Fisher}
G_{F,\lambda\lambda}=\langle\delta\rho \, \delta\rho\rangle^{(\sigma)}_{\lambda\lambda}  = \frac{1}{2} \mbox{tr} \left( \delta\rho \frac{d}{d (\delta\lambda)}\log \left(\sigma + \delta\lambda \delta\rho\right)|_{\delta\lambda = 0}\right).
\ee
This is known as the Fisher information metric in the literature.

On the other hand, a second notion of the quantum distance between the same two states is given by the {\it fidelity susceptibility}, which is defined by
\begin{equation}
\label{fid-sus}
G_{\lambda\lambda} = \partial_{\lambda}^2 F \, ,
\end{equation}
where $F$ is the {\it quantum fidelity}
 defined in terms of the initial and final density matrices $\sigma$ and $\rho$,   
\begin{equation} \label{denmat}
F=\mbox{Tr}\sqrt{\sqrt{\sigma_{\lambda}}\,\rho_{\lambda+\delta\lambda}\sqrt{\sigma_{\lambda}}} \, .
\end{equation}

For classical states when the density matrices commute,  \eqref{Fisher} and \eqref{fid-sus} become equivalent up to an overall numerical factor. Hence, for classical states, the definition for the distance between quantum states is unique.

The first holographic computation of the Fisher information metric \eqref{Fisher} was performed in  \cite{Lashkari:2015hha} using its relation to the second order variation of relative entropy.   On the other hand, the holographic dual of \eqref{denmat} was first proposed in \cite{MIyaji:2015mia}, but only for pure states when 
\eqref{denmat} reduces to an inner product between nearby states,
\begin{equation}\label{buresmet}
|\langle\psi_{\lambda}(x)|\psi_{\lambda+\delta\lambda}(x)\rangle |=1-G^{(\mathrm{pure})}_{\lambda\lambda}(\delta\lambda)^2+\dots \, .
\end{equation}
Here, $G^{(\mathrm{pure})}_{\lambda\lambda}$ now refers to the  fidelity susceptibility for the pure state $|\psi_{\lambda}\rangle$. 
These authors consider the CFT vacuum state $|\psi_\lambda\rangle$ dual to pure AdS and deform it by an \emph{exactly marginal} perturbation to obtain  the state $|\psi_{\lambda+\delta\lambda}\rangle$. In the dual gravity picture, this corresponds to a Janus solution, where the pure AdS is deformed by a dilaton \cite{Bak:2003jk}.  

For a holographic CFT$_d$ on $\mathbb{R}^{d}$ with marginal deformation of dimension $\Delta = d$, in \cite{MIyaji:2015mia} it was shown that\footnote{Note that \cite{MIyaji:2015mia} only considers marginal deformations of the ground state and not general massive deformations which we shall be considering in what follows. However, later on, we will also consider excited mixed states due to scalar perturbations and discuss the importance of marginal perturbations. 
}
\begin{equation}\label{regulatedfisher}
G^{(\mathrm{pure})}_{\lambda\lambda}=\frac{n_d}{G}\frac{\mbox{Vol}(V_{d-1})}{\epsilon^{d-1}}.
\end{equation}
Here $V_{d-1}$ is the spatial $(d-1)$ dimensional volume at the boundary, $G$ is $d+1$-dimensional Newton's constant and $n_d$ an $O(1)$ constant.\footnote{Here we have written the factor of Newton's constant explicitly. In the boundary field theory computation, it comes from the leading-order term of the boundary two-point function (in the $1/N$ expansion of a large $N$ CFT) via $\frac{L^{d-1}}{G}\propto N^2$ ($L$ is AdS radius). The matching of bulk and boundary computations of the fidelity susceptibility in \cite{MIyaji:2015mia} thus only involves the leading-order contribution in Newton's constant on both sides. In other words, the fidelity susceptibility computed there in terms of dual gravity quantities, is only the leading order semiclassical term of the full boundary fidelity susceptibility.} From the field theory point of view, $n_d/G$ is simply proportional to the central charge $C_T$ of the CFT considered.

An essential new ingredient of our approach in the present paper is to  consider mixed states. We will consider  a mixed state in a  bipartite system, where $\sigma_{R,\lambda}$ and $\rho_{R,\lambda +\delta\lambda}$ denote the reduced density matrices for the subregion $R$ corresponding to a decomposition of the full Hilbert space $H_{\mathrm{full}}=H_R\otimes H_{R^{(c)}}$. 
In the first part of our work, we will consider a holographic dual for the Fisher information metric corresponding to a spherical subregion $R$ in the holographic CFT. On the gravity side, we  consider a regularized volume enclosed by the RT surface, and discuss how it is related to the Fisher information for a mixed state.

For mixed states as defined above, in analogy to \eqref{denmat}, the reduced fidelity may be defined by
\begin{equation} \label{Rdenmat}
F_R=\mbox{Tr}_{R^{(c)}} \sqrt{\sqrt{\sigma_{R,\lambda}}\,\rho_{R, \lambda+\delta\lambda}\sqrt{\sigma_{R, \lambda}}}\,.
\end{equation}
In the above, $\mbox{Tr}_{R^{(c)}}$ denotes partial tracing over the complementary region $R^{(c)}$. Accordingly,  the reduced fidelity susceptibility is given by
\begin{equation}
\label{Rfid-sus}
G_{R,\lambda\lambda} = \partial_{\lambda}^2 F_R \, .
\end{equation}
As long as the reduced density matrices in the vacuum and in the excited state are commuting i.e. simultaneously diagonalizable, \eqref{Rfid-sus} is the same as the Fisher information metric corresponding to those reduced density matrices. This is automatically true when we deal with classical states. Hence for this restricted class of states, our proposal also serves as holographic dual for the reduced fidelity susceptibility. 

In \cite{Alishahiha:2015rta}, in the context of proposing a gravity dual for complexity, Alishahiha proposed a holographic dual of the reduced fidelity susceptibility \eqref{Rfid-sus} in terms of the {\it volume} enclosed by the RT surface $\gamma(R)$. 
However, this quantity is UV divergent, while the Fisher information metric for a general mixed state must be finite. Hence, at least for the class of states for which the two notions of information metric introduced above coincide, this proposal yields contradiction. In contrast, our proposal  predicts a manifestly finite Fisher information metric. In addition,  at least for the above-mentioned restricted class of states, our proposal gives rise to a UV-finite reduced fidelity susceptibility.
 
The essential ingredient of our proposal for the holographic dual $G_{R,mm}$ of the reduced fidelity susceptibility is to consider the difference of two volumes which yields a finite expression,
\begin{gather}
{\cal F} = C_d ( V^{(m^2)} - V^{(0)}) \, . \label{eqF}
\end{gather}
Here the first volume in the bracket on the right-hand side is evaluated for a second-order fluctuation about  AdS space involving the stess-energy tensor, and the second term at zeroth order, i.e.~for AdS space itself. The fluctuation considered is dual to the energy-momentum tensor on the field-theory side. The proposal \eqref{eqF} modifies the pure-state volume expression \eqref{regulatedfisher} in a natural way such as to obtain a finite expression.   $C_d$ is a dimensionless constant which cannot be fixed from first principles on the gravity side.\footnote{A similar undetermined coefficient is also present in the pure state proposal of
\cite{MIyaji:2015mia}, as is seen from  equations (2),  (9) and (18) in that paper.}
We will determine $C_d$ by comparison with results for the relative entropy \cite{Blanco:2013joa,Beach:2016ocq}. For metric and marginal perturbations, this coefficient depends only on the spacetime dimension, while for relevant scalar perturbations also the operator dimension enters.
 
We suggest that the holographic reduced fidelity susceptibility is obtained by taking the second order variation of ${\cal F}$ with respect to $m$, 
\be
\label{fid-sus11}
G_{R,mm} =  \partial^2_m {\cal F}.
\ee 
The definition \eqref{fid-sus11}, along with the proposal \eqref{eqF}, ensures that the holographic fidelity susceptibility is finite, as required for mixed states. $G_{R, mm} $ scales as $G_{R,mm} \propto R^{2d}$, with $R$ the radius of the spherical entangling region in the dual field theory in $d$ spacetime dimensions. This scaling behaviour is expected, as we discuss below.  Moreover, the volume difference in \eqref{eqF} entirely encodes the dependence on the shape of the entangling region.

The  particular scaling behaviour  $G_{R,mm} \propto R^{2d}$ also follows from the alternative definition for Fisher information proposed in \cite{Lashkari:2015hha} in terms of the {\it relative entropy} $\Delta S$, which is a measure of entropic distance between two states. An example is the relative entropy measuring this distance between a perturbed state and the ground state.
The  Fisher information metric proposed in \cite{Lashkari:2015hha}  is given by the  second order variation of $\Delta S$  measuring the entropic distance between the ground state of the boundary CFT and the state  obtained by  perturbing this ground state  by injecting energy. As holographically shown in \cite{Blanco:2013joa}  and later confirmed in \cite{Sarosi:2016oks,Sarosi:2016atx} by a direct field theory computation, the relative entropy for this perturbation scales precisely as $R^{2d}$ for a spherical subregion of radius $R$, at quadratic order in energy fluctuations. In fact as discussed in  \cite{Sarosi:2016atx}, when taking into account a calculational issue, both the expressions of relative entropy obtained holographically in \cite{Blanco:2013joa} and from the field theory computation of \cite{Sarosi:2016atx} match exactly including the prefactor. Consequently, the Fisher information metric also scales as $R^{2d}$. Now for the restricted class of states introduced above, the reduced fidelity susceptibility coincides with the Fisher information metric. Therefore, for this class of states, the reduced fidelity susceptibility $G_{R,mm}$ also scales with $R^{2d}$.

Our proposal of identifying the expression \eqref{fid-sus11} with the holographic dual of the fidelity susceptibility for mixed states, using the renormalized volume proposal \eqref{eqF},  thus provides a finite expression with the correct scaling behaviour $R^{2d}$. This expression encodes all information about the shape of the entangling region.

As a further example, we also consider the fluctuations caused by the insertion of a scalar  in AdS and compute the corresponding subtracted volume at the quadratic order in the perturbation parameter. We obtain a scaling behaviour of the corresponding contribution to the Fisher information metric of the form  $R^{2\Delta}$, where $\Delta$ is the scaling dimension of the operator dual to the scalar bulk AdS field. This behaviour is again consistent with the quadratic variation of the relative entropy for such fluctuations \cite{Blanco:2013joa}, and hence our arguments in support of the conjecture given above apply in this case as well.  This examples thus provides a further support for our holographic proposal for the Fisher information metric.

In the second part of this work, we argue that the leading $1/N$ quantum correction to the Fisher information metric is related to the corresponding quantum correction to the boundary entanglement entropy. According to  \cite{Faulkner:2013ana, Jafferis:2015del}, this boundary entanglement entropy correction in turn coincides with the bulk entanglement entropy. Our proposal thus implies that the leading $1/N$ quantum correction to Fisher information is related to the bulk entanglement entropy.
This proposal is motivated by providing an argument for relating the reduced Fisher information to the {\it canonical energy} as defined in  \cite{Hollands:2012sf}. Then, the connection between $1/N$ quantum correction to canonical energy and the bulk modular Hamiltonian \cite{Jafferis:2015del} justifies our proposal.

The bulk entanglement entropy which is seen as the $1/N$ quantum correction to the boundary entanglement entropy, is hard to obtain directly from a bulk computation for a generic bulk state. Therefore, our proposed duality between the bulk entanglement entropy and the $1/N$ quantum correction to Fisher information can be one first step towards understanding the quantum nature of  the bulk theory. In particular, this connection might play a pivotal role in understanding the Hilbert space structure of quantum gravity in the bulk. Consequently, one might further expect this connection to shed some light on the reconstruction of local bulk fields from boundary CFT operators beyond the semiclassical limit. 

Our paper is organised as follows. In section \ref{sec:hol-fid} and section \ref{sec:conjec} we establish the two proposals mentioned above, namely (A) we discuss a holographic quantity which is associated to the Fisher information metric and (B) we propose a connection between the leading $1/N$ quantum correction to  the Fisher information metric and the bulk entanglement entropy. We conclude in section \ref{sec:conclusion} and discuss some of the consequences of our proposals, as well as a physical consistency check. We also discuss directions for future work.

\section{Proposing a holographic dual for Fisher information metric}\label{sec:hol-fid}

For the holographic dual, we consider an asymptotically AdS
spacetime using Fefferman-Graham coordinates.  For the boundary CFT, this
amounts to considering states whose density matrix deviates
perturbatively from that of the vacuum state, with the change in the
boundary stress tensor playing the role of the perturbation
parameter. For this excited state, we  compute the volume under the
RT surface corresponding to a spherical entangling region at the
boundary. This volume is generally UV divergent. However, subtracting
the RT volume for the same spherical subregion in the vacuum state
yields a finite result.  We propose that the Fisher
information metric is given by the second order variation of
this regularized volume with respect to the perturbation parameter. In what follows we
will consider $d>2$.\footnote{The $d=2$ case is special in
  the sense that the perturbative expansion of the regularized volume
  does not contain a quadratic term. The reason is that for $d=2$, $h_2(z)$ in \eqref{h-z} vanishes identically and there is no contribution towards the minimal area surface at the second order in $m R^d$.}
  
 \subsection{Stress-tensor perturbations}
To elaborate, let us consider a perturbation of $AdS_{d+1}$ given in
Fefferman-Graham coordinates of the form
\be
\label{metric}
ds^2 = \frac{L^2}{z^2} \left[f(z) dz^2 + \frac{1}{f(z)} dt^2 + d\rho^2 + \rho^2 d\Omega_{d-2}^2\right],
\ee 
where 
\begin{equation} \label{eqT} 
f(z) = 1+ m z^d \, .
\end{equation}
$L$ is the radius of the AdS space-time.

In order to find the minimal RT surface in this perturbed AdS
spacetime corresponding to a ball-shaped entangling region of radius
$R$ at the boundary, we proceed by parametrizing the RT surface as $\rho = h(z)$. Then, on the $t = 0$ slice the RT area functional takes the form
\be
\label{area1}
A = L^{d-1} \Omega_{d-2} \int_{\epsilon}^{R_t} \frac{dz}{z^{d-1}} \left(h(z)\right)^{d-2}\sqrt{f(z) + \left(h'(z)\right)^2} ,
\ee 
where $\Omega_{d-2}$ is the volume of the unit $(d-2)$ sphere, given by
\begin{equation}
\Omega_{d-2} = 2 \,
\frac{\pi^{\frac{d-1}{2}}}{\Gamma{\left(\frac{d-1}{2}\right)}} \, .
\end{equation}
$R_t$ is the turning point of the bulk minimal surface. 

In order to find the minimal surface, we have to minimize the area
functional \eqref{area1} to solve for $h(z)$. It is hard to solve the
equations of motion analytically. We therefore aim at solving them
perturbatively in orders of $m R^d \ll 1$ and look for a  solution of the form (up to linear order; quadratic order to be done later)
\be
\label{h-z}
h (z) = h_0 (z) + m h_1(z) \, .
\ee
As shown in \cite{Jyoti-Taka}, this gives
\begin{align}
\label{h0h1}
h_0 &= \sqrt{R_t^2 - z^2} \, , \nonumber \\
h_1 &= \frac{2 R_t^{d+2} - z^d (R_t^2 + z^2)} {2(d+1) \sqrt{R_t^2 -
    z^2}} \, .
\end{align}
With these ingredients, we now move on to compute the volume under the
RT minimal surface in the bulk. After performing the integrations over
the boundary coordinates $\rho$ and $\Omega$, this is given by
\be
\label{V-RT1}
V_{RT} = \frac{L^{d} \Omega_{d-2}}{d-1} \int_\epsilon^{R_t} \frac{dz}{z^{d}} \left(h(z)\right)^{d-1} \sqrt{f(z)}.
\ee 
Our aim is now to compute the variation of this volume order by order
in the perturbation $m R^d$. 

\subsubsection{At linear order in the stress-energy tensor}

From the fundamentals of AdS/CFT duality, we know that there is a relation between such Fefferman-Graham type expansions of AdS metric and the corresponding expectation values of boundary stress tensor \cite{Skenderis:2002wp}. In order to find the leading variation in the RT volume we first expand \eqref{h-z} up to leading order in 
$m R^d$,
\be
\label{h-z-approx}
h(z)  \approx \sqrt{R^2 - z^2} \left(1 - m  \frac{z^d (R^2 + z^2)} {2(d+1) (R^2 - z^2)}\right),
\ee

Inserting \eqref{h-z-approx} into \eqref{V-RT1} and expanding
individual terms in the integral again, we have
\begin{align}
\label{V-RT1-approx}
V^{(m)}_{RT} &\approx \frac{L^{d} \Omega_{d-2}}{d-1} \int_\epsilon^{R_t} dz \frac{(R^2 - z^2)^{\frac{d-1}{2}}}{z^{d}} \nonumber \\
&\times \left(1 - m \frac{(d-1) z^d (R^2 + z^2)} {2(d+1) (R^2 - z^2)}\right) \left(1+ \frac{m z^d}{2}\right).
\end{align}
Here the superscript $m$ signifies that this is the volume
under the RT surface corresponding to a perturbed geometry. So as a next step, in order to find the linear variation in $m$, we
subtract from it the same volume for the unperturbed background of pure AdS obtained by setting $m=0$ in \eqref{metric}. This yields
\begin{align}
V^{(m)}_{RT}  - V^{(0)}_{RT}  &\approx m \frac{L^{d} \Omega_{d-2}}{(d-1)(d+1)} \int_\epsilon^{R}
 dz (R^2 - z^2)^{\frac{d-3}{2}} \left(R^2 - d z^2\right) \nonumber \\
&= 0,
\end{align}
where in the first line, we have only kept terms up to order $m R^d$
in the integrand. Furthermore, we have replaced $R_t$ by $R$ since the
term linear in $m R^d$ in $R_t$ gives a quadratic correction to the volume.
This shows that the leading correction to the volume under RT surface
vanishes identically as claimed in \cite{Alishahiha:2015rta,
  Ben-Ami:2016qex}. The vanishing of this linear term in $m$ is also
crucial for our proposal (A) to work, as will be seen in the next section.

\subsubsection{At quadratic order in the stress-energy tensor} \label{quad-T}
At quadratic order, we have 
\begin{equation}
\label{pert-grav}
f(z) = 1+ m z^d + \frac{1}{4} m^2 z^{2d} \,
\end{equation}
in place of \eqref{eqT}, 
where  the coefficients of individual terms in the expansion is fixed
by comparing with the Fefferman-Graham expansion of AdS black hole.
Now in order to compute the quadratic ${\cal O} (m^2)$ correction to the RT surface in the bulk, we start with an ansatz
\begin{equation}
\label{h-z-quad}
h (z) = h_0 (z) + m h_1(z) + \frac{m^2}{\sqrt{R_t^2-z^2}} h_2(z) \, ,
\end{equation}
with $h_0$ and $h_1$ as given in \eqref{h0h1}.

The equation of minimal surface for $h_2$ is again obtained by extremizing the area functional \eqref{area1} ,
\begin{equation}
\label{eomh2}
h_2''(z) + \frac{(d-1) R_t^2}{z \left(z^2-R_t^2\right)}  h_2'(z) + C_d(z) =0,
\end{equation}
with $C_d(z)$ being a complicated function of $z$. 
It is very hard to solve \eqref{eomh2} for general dimensions. However, 
as an illustration, we will consider  $d=3$ when \eqref{eomh2}
can be readily solved to yield
\begin{align}
\label{h2sol}
h_2(z) &= \frac{1}{320}\Bigg(\left(160 \, c_1 R_t-11 R_t^8\right) \log (z-R_t)\nonumber\\
&+\left(59 R_t^8-160 c_1 R_t\right) \log (R_t+z)+320 c_1 z-\frac{20 R_t^9}{R_t+z}\nonumber\\
&-90 R_t^7 z+34 R_t^6 z^2-30 R_t^5 z^3+22 R_t^4 z^4-\frac{9 R_t^2
  z^6}{2}\Bigg)+c_2 \, .
\end{align}
$c_1$ and $c_2$ are integration constants which should be suitably
chosen in order to extract the physical solution. We note from the
solution that in order to ensure \linebreak ${h_2(z)}/{\sqrt{R_t^2-z^2}}
\rightarrow 0$ as $z \rightarrow R_t$, we must have 
\begin{equation}
c_1 =\frac{11}{160} R_t^7 \quad \mathrm{and} \quad  c_2 =\frac{1}{640} R_t^8 \left(113 -96 \log
  (2 R_t)\right).
  \end{equation}
  
 Consequently, the turning point also receives a new
contribution at this order of perturbation theory and in terms of the radius of the entangling region $R$ is given by
\be
R_t=\frac{3}{640} m^2 R^7 (29+32 \log (2))-\frac{m R^4}{4}+R \, .
\ee

Now expanding \eqref{V-RT1} up to quadratic order, we find
\be
\label{V-RT2-approx-d}
V^{(m^2)}_{RT} - V^{(0)}_{RT}\approx \frac{L^{d} \Omega_{d-2}}{d-1} {\cal{A}}_d \, m^2 R^{2d},
\ee 
where in general dimensions, ${\cal{A}}_d$ is an involved constant depending on $d$ which we do not write out explicitly.
In particular, for $d=3$, \eqref{V-RT2-approx-d} takes the form
\be
\label{V-RT2-approx}
V^{(m^2)}_{RT} - V^{(0)}_{RT}\approx \frac{21\pi L^3 R^6m^2}{128}.
\ee

This is the first central result of our paper. We see that we have
arrived at a UV-finite notion of a regularized volume under the RT
surface. It is of second order in $m$. We will exploit this fact for
proposing it as the
holographic dual of  Fisher information. We emphasize that the
finiteness of the regularized volume defined here is critical for this
proposal. In particular, this ensures a meaningful gravity dual for
mixed states.\footnote{See also \cite{Caceres:2016xjz} and \cite{Couch:2016exn} for some recent suggestions on possible regularized quantities which could be related to complexity.}

We are thus lead to propose that the holographic dual of the Fisher
information metric is given by
\begin{equation}
\label{Rfid-sus-hol}
G_{F,mm} = C_d \partial_{m}^2 {\cal{F}},
\end{equation}
with
\be
\label{def-hol-fid}
{\cal{F}} = \frac{\pi^{\frac{3}{2}} d
  \left(d-1\right)\Gamma\left(d-1\right)}{G 2^{d+1} (d+1)
  \Gamma\left(d+\frac{3}{2}\right)  L {\cal{A}}_d}
\left(V^{(m^2)}_{RT} - V^{(0)}_{RT}\right) \, .
\ee
Inserting \eqref{def-hol-fid} back into \eqref{Rfid-sus-hol} yields
\be
\label{holFisher1}
G_{F,mm} = \partial_{m}^2 {\cal{F}} = \frac{\pi^{\frac{3}{2}} d
  \Omega_{d-2} \Gamma\left(d-1\right)}{G 2^{d+1} (d+1)
  \Gamma\left(d+\frac{3}{2}\right)} L^{d-1} R^{2d} \, .
\ee
As mentioned before, $m R^d$ plays the role of
perturbation parameter in the dual bulk picture, in agreement with the
holographic dictionary. The prefactor in \eqref{def-hol-fid} is chosen in such a way as to
ensure coincidence with the result for relative entropy given in
\cite{Blanco:2013joa,Lashkari:2015hha}. We discuss the motivation for this matching
below in section 2.3. Here we stress that the result \eqref{holFisher1}
has three essential properties: First, it provides a finite result for
mixed states as required; second, it reproduces the correct scaling
with $R$; third, the shape of the entangling region enters only
through the volume.

Note that so far, \eqref{Rfid-sus-hol} applies only to ball-shaped
regions in the CFT.  One may also wish to consider general entangling
regions, e.g.~strips. In such cases the $O(m R^d)$ terms do not
necessarily vanish \cite{Ben-Ami:2016qex}. However, in principle, one
can still define the holographic dual to Fisher information metric as
the second order variation of the regularized RT volume with respect
to the mass parameter, with this parameter playing the role of perturbation parameter in the dual bulk theory.

\subsection{Scalar perturbations }\label{subsec:scalerpert}

So far we considered  perturbations arising from  stress-energy tensor
deformations of the ground state. Here we turn to the question whether
our proposal is applicable to other non-trivial states, e.g.~states
that are deformed from pure AdS due to some matter perturbation\footnote{We are grateful to Nina Miekley for collaborating on the results presented  in this subsection.}. This
will then provide further support for  our proposal. Here we consider
the case that the boundary state is perturbed by a scalar operator
$\mathcal{O}_{\Delta}$ of conformal dimension $\Delta$, and show that
our proposal for the Fisher metric for mixed states holds in this case too.
We turn to scalar perturbations of the type studied in
\cite{Blanco:2013joa}, where  a gravity calculation of relative
entropy is provided. The bulk dual of such perturbations correspond to  a scalar field
\be
\label{bc}
\phi = \gamma \mathcal{O}_{\Delta} z^{\Delta}, 
\ee
backreacting on the background geometry. $\gamma$ is a normalization constant. The generic perturbations to the linear order in boundary stress tensor and quadratic order in $\mathcal{O}_{\Delta}$ take the  form \cite{Blanco:2013joa}
\be 
\delta g_{\mu\nu}=a
z^d\sum_{n=0}z^{2n}T_{\mu\nu}^{(n)}+z^{2\Delta}\sum_{n=0}z^{2n}\sigma_{\mu\nu}^{(n)}+\dots
\, ,
\ee
with $n$ denoting  the $2n$ derivatives appearing in the corresponding term and $a=\frac{2}{d}\frac{G}{L^{d-1}}$. 
The leading order $n = 0$ term in this derivative expansion is given by
\be
\label{sigma}
\sigma_{\mu\nu}^{(0)}=-\frac{\gamma^2}{4(d-1)}\eta_{\mu\nu}\mathcal{O}_{\Delta}^2 \equiv -\frac{1}{4}\gamma_0^2\mathcal{O}_{\Delta}^2 \eta_{\mu\nu},
\ee
with $\gamma$ being the same dimensionless normalization constant as
in \eqref{bc}. 

In the previous section, we already considered perturbations due to
the stress-energy tensor up to quadratic order. Therefore, in what
follows, we will only focus on the case where the bulk perturbation is
due to a scalar field, which means that  we only consider the contribution of the second term in \eqref{sigma}. Redefining the scalar condensate as $\tilde{\epsilon}^2 = -
\mathcal{O}_{\Delta}^2 \gamma_0^2$, the metric perturbation takes the form 
\be
\label{pert-scalar}
\delta g_{\mu\nu}=\frac{1}{4} \tilde{\epsilon}^2 z^{2 \Delta} \eta_{\mu\nu}
\, .
\ee
Now in order to compute correctios to the bulk RT surface  up to
quadratic ${\cal O} ({\tilde \epsilon}^2)$, in a  spirit similar to \eqref{h-z-quad}, we begin with the ansatz
\begin{equation}
\label{h-zscal1}
h (z) = h_0 (z) +  {\tilde \epsilon} h_1(z) + \frac{{\tilde \epsilon}^2}{\sqrt{R_t^2-z^2}} h_2(z) \, ,
\end{equation}
However unlike the case of stress-tensor perturbation,  now there is no linear contribution to the perturbation, i.e $ h_1(z)  = 0 $. This is a consequence of the form of perturbation given in \eqref{pert-scalar}. $h_0(z)$ is the same as in \eqref{h0h1}, i.e, 
\be
h_0 (z) = \sqrt{R_t^2 - z^2} \, .
\ee
The equation of minimal surface for $h_2$ is again obtained by extremizing the area functional \eqref{area1} with
\begin{equation}
\label{pert-grav2}
f(z) = 1+ \frac{1}{4} {\tilde\epsilon}^2 z^{2\Delta} \, .
\end{equation}
This gives
\begin{equation}
\label{eomh2scalar}
h_2''(z) + \frac{(d-1) R_t^2} {z \left(z^2-R_t^2\right)} h_2'(z) +\frac{z^{2 \Delta } \left(R_t^2 (d-\Delta
   -2)+(\Delta +1) z^2\right)}{4 \left(z^2-R_t^2\right)} = 0.
\end{equation}
\eqref{eomh2scalar} can be readily solved to yield 
\bea
h_2(z) &=& \frac{1}{8} \left[-\frac{(-d+\Delta +2) z^{2 \Delta +2} \, _3F_2\left(1,\Delta + \frac{1}{2},\Delta +1;\Delta +2,-\frac{d}{2}+\Delta
+2;\frac{z^2}{R_t^2}\right)}{(\Delta +1) (-d+2 \Delta +2)} \right. \nonumber \\
 &+& \left.  \frac{(\Delta +1) z^{2 \Delta +4} \, _3F_2\left(1,\Delta +\frac{3}{2},\Delta +2;\Delta
   +3,-\frac{d}{2}+\Delta +3;\frac{z^2}{R_t^2}\right)}{(\Delta +2) (-d+2 \Delta +4)
   R_t^2} \right. \nonumber \\ 
  &+& \left.  8 \left(C_1 R_t \left(\frac{z}{R_t}\right)^d \,
   _2F_1\left(\frac{d-1}{2},\frac{d}{2};\frac{d+2}{2};\frac{z^2}{R_t^2}\right)+C_2\right)\right],
\eea
$C_1$ and $C_2$ being the integration constants which can be fixed by demanding that \linebreak ${h_2(z)}/{\sqrt{R_t^2-z^2}}
\rightarrow 0$ as $z \rightarrow R_t$. While this is is hard to
implement in general dimensions, this is straightforward for the cases
$d=3$ and
$d=4$. Here we shall concentrate on $d=3$ and integer values of
$\Delta > 1$. We have also checked the results for higher dimensions
$d$ for integer values of $\Delta > \frac{d-2}{2}$. 

For $d=3$, $C_1$ and $C_2$ take the forms
\begin{gather}
C_1 = -\frac{\Delta  R_t^{2 \Delta +1}}{6 (2 \Delta -1) (2 \Delta +1)}
\, , \qquad 
C_2 =   -\frac{\Delta \,  \Gamma\left(\Delta -\frac{1}{2}\right)
  R_t^{2 \Delta +1}}{24 \,  \Gamma\left(\Delta +\frac{3}{2}\right)} \, .
\end{gather}
Furthermore, the turning point also receives a correction of the form
\be
R_t = R - {\tilde\epsilon} ^2 C_2  R^{2 \Delta +1} R_t^{-2 \Delta -2}
\, .
\ee
Expanding \eqref{V-RT1} up to quadratic order in ${\tilde \epsilon}$, we obtain the difference in volume as 
\be
\label{V-RT2-approx-scalar}
V^{({\tilde \epsilon}^2)}_{RT} - V^{(0)}_{RT}\approx \frac{\pi L^3\tilde{\epsilon}^2}{16} R^{2 \Delta } \left(\frac{1}{(\Delta -1) \Delta }-\frac{2 \Gamma (\Delta
   )}{\Gamma (\Delta +1)}+\frac{2 \Delta  (\log 16 - 2) + 4 \Delta  H_{\Delta }-2}{4
   \Delta ^2-1}\right),
\ee 
where $H_{\Delta }$ is the harmonic number of order $\Delta$.
This result generalizes to general dimensions $d$, where it becomes
\be
\label{V-RT2-approx-scalar-d}
V^{(\tilde{\epsilon}^2)}_{RT} - V^{(0)}_{RT}\approx \frac{L^{d} \Omega_{d-2}}{d-1}
{\cal{B}}_{d,\Delta} \, {\tilde\epsilon}^2 R^{2\Delta} \, .
\ee 
Here, ${\cal{B}}_{d,\Delta}$ is a complicated dimension-dependent constant which for $d=3$ can be read-off from \eqref{V-RT2-approx-scalar}.

Thus, in general dimensions and for scalar perturbations, our proposal
for the corresponding entries in  Fisher information metric in terms of the regulated volume is given by
\begin{equation}
\label{Rfid-sus-hol2-scalar}
G_{F,\tilde{\epsilon}\tilde{\epsilon}} = \partial_{\tilde{\epsilon}}^2
{\cal{F}} \, , \qquad {\cal{F}} = C_{d, \Delta} (V^{(\tilde{\epsilon}^2)}_{RT} -
V^{(0)}_{RT}) \, ,
\end{equation}
in analogy to \eqref{Rfid-sus-hol}, with
\be
\label{def-hol-fid2-scalar}
{\cal{F}} = \frac{\pi^{\frac{3}{2}}
  \left(d-1\right) \left(\Delta - \frac{(d-2)^2}{2(d-1)}\right)\Gamma\left(\Delta - \frac{d}{2} +1\right)}{8 G 
  \Gamma\left(\Delta - \frac{d}{2}+\frac{5}{2}\right)  L {\cal{B}}_{d,\Delta}}
\left(V^{(\tilde{\epsilon}^2)}_{RT} - V^{(0)}_{RT}\right) \, .
\ee
From \eqref{def-hol-fid2-scalar} and \eqref{Rfid-sus-hol2-scalar}, we  obtain 
\be
\label{holFisher1-2-scalar}
G_{F,\tilde{\epsilon}\tilde{\epsilon}} = \partial_{\tilde{\epsilon}}^2 {\cal{F}} = \frac{\pi^{\frac{3}{2}}
  \left(d-1\right) \left(\Delta - \frac{(d-2)^2}{2(d-1)}\right)\Gamma\left(\Delta - \frac{d}{2} +1\right)}{8 G 
  \Gamma\left(\Delta - \frac{d}{2}+\frac{5}{2}\right) } L^{d-1} \Omega_{d-2} R^{2\Delta}.
\ee
This is similar to what we obtained for the  correction
quadratic in the stress-energy tensor in \eqref{holFisher1}. Here
however, the entry into the Fisher information metric corresponds to a
perturbation in  a new state parameter  $\tilde \epsilon$ instead of
$m$. Again we obtain a finite result with the
expected scaling with $R$, independent of the shape of the entangling
region.  We have chosen the prefactor $C_{d, \Delta}$ in
\eqref{Rfid-sus-hol2-scalar} in such a way that the result for ${\cal F}$ 
  coincides with the relative entropy in the presence of scalars given
  in \cite{Blanco:2013joa,Beach:2016ocq}.  For marginal perturbations for which
  $\Delta = d$, the coefficient $C_{d, \Delta}$ depends only on the
  spacetime dimension, while for relevant perturbations in particular
  it depends on the operator dimension $\Delta$ as well.

\subsection{Connection to canonical energy and boundary relative entropy}

In a related development, \cite{Lashkari:2015hha} connects
the quantum Fisher information corresponding to perturbations
of the CFT vacuum density matrix of a ball-shaped region to the canonical energy for perturbations in the corresponding Rindler wedge of the dual $AdS$ space-time. This is obtained by using the definition of boundary relative entropy
\begin{eqnarray}
\label{rel-entropy}
S_{\mathrm{rel}}^{(\mathrm{bdy})}(\rho_{\lambda'}||\sigma_{\lambda}) &=& \mbox{Tr} \left(\rho \log \rho \right) - \mbox{Tr} \left(\rho \log \sigma \right) \nonumber \\
&=& \langle\log\rho\rangle_{\rho}-\langle\log\sigma\rangle_{\rho},
\end{eqnarray}
which gives
\be
\label{rel-entropy1}
S_{\mathrm{rel}}^{(\mathrm{bdy})}(\rho_{\lambda'}||\sigma_{\lambda})= \Delta\langle \mathcal{H}^{(\sigma)}_{R}\rangle-\Delta S_{EE}.
\ee
The first term on the right-hand side denotes the change in the expectation value of the modular Hamiltonian $\mathcal{H}_{R}$ corresponding to the change in the reduced density matrix. The modular Hamiltonian corresponding to a reduced density matrix $\sigma_{\lambda}$  is defined through
\be
\label{mod-Ham}
\sigma_{R,\lambda}=\frac{e^{-\mathcal H_{R,\lambda}}}{\mbox{Tr}(e^{-\mathcal H_{R,\lambda}})}.
\ee
Here, the second term represents the change in entanglement entropies for the two above-mentioned states.
When the two states in question are perturbatively close to one another, expanding the density matrix $\rho_{\lambda'}$ around $\lambda=0$ in \eqref{rel-entropy1} gives (we have dropped the superscript (bdy) from the left side of \eqref{rel-entropy1} to avoid clutter)
\be
\label{fisher-rel1}
G_{F,\lambda\lambda}=\langle\delta\rho \, \delta\rho\rangle^{(\sigma)}_{\lambda\lambda} = \frac{\partial^2}{\partial\lambda^2}S_{\mathrm{rel}}(\rho_\lambda||\rho_0),
\ee
where the left-hand side denotes the Fisher information metric as defined in \eqref{Fisher}. $\rho_0 = \sigma$ is identified with the CFT vacuum. 

Furthermore, the right-hand side of \eqref{fisher-rel1} is equal to
the \emph{classical} canonical energy in gravity as defined in \cite{Hollands:2012sf}, i.e, 
\be
\label{result-Lashkari}
\frac{\partial^2}{\partial\lambda^2}S_{\mathrm{rel}}(\rho_\lambda||\rho_0)\Big{|}_{\lambda=0} = {\cal E}-2 \int_{\Sigma}\xi^\mu\frac{\partial^2E^{g}_{\mu\nu}}{\partial\lambda^2}v^\nu.
\ee

All quantities on the right-hand side of \eqref{result-Lashkari}
belong to the gravity side of the correspondence.  $\cal E$ is the classical canonical energy for the unperturbed vacuum state and can be expressed as an integral of boundary stress-energy tensor \cite{Hollands:2012sf}, 
\be
\label{can-en}
{\cal E} = \int_{\Sigma}\xi^\mu \, T_{\mu\nu} \, {v}^\nu,
\ee
where $\Sigma$ is any Cauchy slice in the entanglement wedge corresponding to the ball-shaped entangling region in the boundary. $\xi$ is the conformal Killing vector, $v$ is the volume form defined as (for a $D$-dimensional spacetime)
\[
v_{\nu}=\frac{1}{D!}\sqrt{\tilde{g}}\,\epsilon_{\nu, \mu_1\mu_2\dots\mu_D}dx^{\mu_1}\wedge dx^{\mu_2}\dots\wedge dx^{\mu_D},
\]
and $\epsilon$ being the usual Levi-Civita tensor.  $E^g$ denotes gravitational equations of motion with proper cosmological constant, e.g, for pure gravity in AdS 
\be
\label{einstein}
E^{g}_{\mu\nu}=\frac{1}{16\pi G}\sqrt{-g}\left(R_{\mu\nu}-\frac{1}{2}Rg_{\mu\nu}+\frac{1}{2}\Lambda g_{\mu\nu}\right).
\ee

For the perturbed $AdS_{d+1}$ space-time as in \eqref{metric}, and for the
case when the entangling region is a sphere, the canonical energy as
on the right-hand side of \eqref{result-Lashkari} may be computed
explicitly. 
One can also independently compute the second order variation of the
relative entropy,
$\frac{\partial^2}{\partial\lambda^2}S_{\mathrm{rel}}(\rho_\lambda||\rho_0)$. Both calculations were done for $d=2$ in \cite{Lashkari:2015hha} and were shown to match explicitly.  
In the holographic setup, $\lambda$ is again identified with the
boundary energy parametrized by $m, $ which appears as a mass parameter in \eqref{metric}. 
In general dimensions, the second variation of the relative entropy reads \cite{Blanco:2013joa}
\begin{align}
\label{holFisherd}
\frac{\partial^2}{\partial\lambda^2}S_{\mathrm{rel}}(\rho_\lambda||\rho_0) &= \frac{\pi^{\frac{3}{2}} d \Omega_{d-2} \Gamma\left(d-1\right)}{G 2^{d+1} (d+1) \Gamma\left(d+\frac{3}{2}\right)} L^{d-1} R^{2d} \nonumber \\
&= G_{F,\lambda\lambda},
\end{align}
where in the last line we have used the definition \eqref{fisher-rel1}.
The basic ingredients in this computation is \eqref{rel-entropy1} and the fact that in this particular
case of spherical entangling region in CFT, the modular Hamiltonian has a local expression in terms of the boundary stress energy tensor as  
\be
\label{mod-ham2}
\mathcal{H}_{R}=\int_{|x| < R} d^{d-1} x \frac{R^2 - |x|^2}{2 R} T_{00}\,, \nonumber \\
\ee
with $T_{00}$ being the temporal component of the stress-energy tensor in the boundary CFT. Hence one can vary both the terms in the right hand side of \eqref{rel-entropy1} up to second order in $m R^d$ which yields \eqref{holFisherd}.

A similar conclusion can be drawn for the deformation with scalar
condensate presented in section \ref{subsec:scalerpert} by noting that
\eqref{holFisher1-2-scalar} is given by the second order variation of
relative entropy with respect to the state parameter $\tilde\epsilon$
defined in \eqref{pert-grav2}. The final expression
\eqref{holFisher1-2-scalar} matches with the expression for the
quadratic variation of the relative entropy with scalar perturbations,
as given
in \cite{Blanco:2013joa}. Moreover, as we also point out later, \eqref{holFisher1-2-scalar} is precisely the behavior that we expect for bulk canonical energy due to scalar perturbations \cite{Beach:2016ocq}. {This extends the validity of our proposal
  to states created by more general perturbations than those
  involving the stress-energy tensor}.

\subsection{Justification from field theory}

It is worth mentioning at this point that \eqref{holFisherd} can be
independently obtained from a computation entirely in field theory
without referring to dual gravity background. This was first done in
\cite{Sarosi:2016oks} for $d=2$ and was generalized to arbitrary
dimensions in \cite{Sarosi:2016atx}. In order to compute the relative
entropy $S_{\mathrm{rel}}(\rho||\sigma)$, these authors first employed
a replica trick as in \cite{Calabrese:2009qy}. The relative entropy can be obtained as a limit from the resulting replicated geometry  \cite{Lashkari:2014yva, Lashkari:2015dia} as 
\be
\label{srel11}
S_{\mathrm{rel}}(\rho||\sigma) = \lim_{n\rightarrow 1} \frac{1}{n-1} \left(\log \tr \rho^n - \log \tr \rho \sigma^{n-1}\right), 
\ee
where
\be
\label{sn1}
S_n =\log \tr \rho^n - \log \tr \rho \sigma^{n-1}.
\ee
Individual contributions to $S_n$ can be obtained by constructing path integrals in the $n$-cover manifold corresponding to the replicated geometry. In order to technically achieve this one needs to go from the CFT on branched cylinder, $R \times S^{d-1}$ to a CFT on the covering manifold $\rm{S}_n \times H^{d-1}$, $H^{d-1}$ being $(d-1)$ - dimensional hyperbolic space and $\rm{S}_n$, a $2 \pi n$ periodic circle. This conformal mapping can be thought of as combining two conformal maps - first a map from $R \times S^{d-1}$ to a branched sphere $\rm{S}_n^d$, and then the second map from $\rm{S}_n^d$ to the covering manifold, $\Sigma_n = \rm{S}_n \times H^{d-1}$.

Finally using the state-operator map the first term in $S_n$ can be written as 
\be
\tr \rho^n = {\cal{N}}_n  \frac{\langle \prod_{k=0}^{n-1} P(\tau_k) P({\tilde\tau}_k) \rangle_{\Sigma_n}}{\prod_{k=0}^{n-1} \langle P(\tau_k) P({\tilde\tau}_k)\rangle_{\Sigma_1}}.
\ee
Here the points, $\tau_k$ and and ${\tilde\tau}_k$ corresponds to, the $t = \infty$ and $t = -\infty$ of the $k$-th Riemann sheet and $P(\tau_k)$ ($P({\tilde\tau}_k)$)  denotes local operator insertion at point $\tau_k$ (${\tilde\tau}_k$) corresponding to the state with reduced density matrix $\rho$. ${\cal{N}}_n$ is the normalization constant. 
A similar expression can be obtained for the second term in the expression for $S_n$ in \eqref{sn1}\footnote{It turns out that the normalization constant ${\cal{N}}_n$ is the same for both the terms and therefore we can set it to $1$ without loss of generality}. 

An operator product expansion for the fields on the $n$-covering manifold
is then substituted in the trace expressions given above. Restricting
to the OPE contribution coming from the stress-energy tensor exchange,
namely the identity Virasoro block, we have\footnote{Furthermore, here it is assumed that the anomaly term is zero, which indeed is the case for odd $d$. If the anomaly term is present, $T_{MN}$ in \eqref{OPE} needs to be redefined as $T_{MN}(\tau_k) - \langle T_{MN}(\tau_k) \rangle_{\Sigma_n}$.}
\be
\label{OPE}
P(\tau_k) P({\tilde\tau}_k) = \langle P(\tau_k) P({\tilde\tau}_k)
\rangle_{\Sigma_n} \left[ 1 + C_{PP}^{MN} \left(\Sigma_n : (\tau -
    \tilde \tau)\right)T_{MN}(\tau_k) \right] .
\ee
We note again that the stress tensor is holographically dual to the metric
perturbation introduced in \eqref{eqT}, which justifies the
restriction to the stress-tensor OPE contribution.
In the case when the size of the subsystem is small, i.e.~when the
radius $R$ of the spherical subregion satisfies $R \ll  1$, inserting
the OPE  contribution into the traces given above results in a
systematic and convergent expansion in $R^{d}$. At leading order, the relative entropy $S_{\mathrm{rel}}(\rho||\sigma)$ scales as
\be \label{sarosi-eq}
S_{\mathrm{rel}}(\rho||\sigma) \propto \epsilon^2 R^{2d}, 
\ee
where $\epsilon = \langle P | T_{tt} | P\rangle$ is the energy of the
system on the cylinder.   Using the explicit proportionality constant
in \eqref{sarosi-eq}  and the relation between $\epsilon$ and the mass
parameter of the black brane, the authors of \cite{Sarosi:2016atx}
showed an exact agreement with the holographic expression for relative
entropy of \cite{Blanco:2013joa}. Their result
 provides  further justification for \eqref{holFisherd}.

Hence, not only do we find our proposal \eqref{holFisher1} to be fully consistent with the expression for Fisher information metric obtained in \eqref{holFisherd}, we now also see that it correctly matches with a direct field theory calculation as mentioned in the previous paragraph. Moreover, in the field theory calculation of relative entropy \eqref{sarosi-eq} and subsequently of Fisher information, the coefficient appearing before the crucial $R^{2d}$ behavior agrees with the gravity calculation of Fisher information as in \eqref{holFisherd}. This also then justifies our choice of prefactors in \eqref{def-hol-fid} and makes our proposal consistent with the Fisher information calculations from both sides of holography.

As discussed before, it is worth mentioning again at this point that for a restricted class of states when 
the reduced density matrices in the vacuum and in the excited state are simultaneously diagonalizable - or in other words, when the subregions are maximally entangled even after perturbation, 
one should expect, analogous to  \eqref{fisher-rel1}, 
\be
\label{fid-rel1}
G_{R, \lambda\lambda} = \frac{\partial^2}{\partial\lambda^2}S_{\mathrm{rel}}(\rho_\lambda||\rho_0),
\ee
$G_{R, \lambda\lambda}$ being the reduced fidelity susceptibility defined in \eqref{Rfid-sus}. For these states, our proposal \eqref{holFisher1} also serves as a holographic dual to reduced fidelity susceptibility while \eqref{def-hol-fid} can be interpreted as the holographic dual to reduced fidelity.

Let us then briefly summarize our results of this section. We show that there is a well-defined, finite notion of regularized volume which serves as the holographic dual to  Fisher information for two perturbatively close states. Both of them are in turn related to the classical canonical energy in the subregion. This set of connections will play a crucial role for the next part of our paper where we make statements regarding their quantum counterpart. Here we also noted that for the special class of states, all the above definitions coincide with the definition of reduced fidelity susceptibility, thus modifying the previously existing proposal of \cite{Alishahiha:2015rta}.

\section{Fisher information and bulk entanglement}\label{sec:conjec}

We now turn to the second part of our proposal on relating the $1/N$ quantum correction to reduced fidelity susceptibility with bulk entanglement entropy. 

\subsection{Bulk entanglement entropy and quantum canonical energy}\label{sec:finalrev}

\begin{figure}
\begin{center}
\includegraphics[width=0.5\textwidth, height=0.28\textheight]{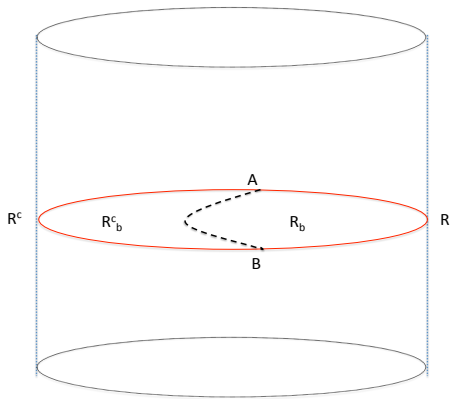}
\end{center}
\caption{At the boundary CFT$_d$ of the global AdS$_{d+1}$ cylinder, we have a disc shaped region $R$ denoted by $AB$ (red line, color online). The dashed (black) line $\gamma$ represents the RT surface which divides the bulk region into two subregions $R_b$ and $R_b^c$. The area of this minimal surface gives the leading semiclassical term of the total boundary entanglement entropy $S_{EE}$. The $\mathcal{O}(G^0)$ term of bulk entanglement entropy of the region $R_b$ is a measure of the first-order quantum correction term $S_{EE,q}$ of $S_{EE}$.\label{genfig}}
\end{figure}

Our investigation of bulk entanglement entropy is motivated by a recent study in \cite{Faulkner:2013ana}. There the authors argue that the $1/N$ quantum correction to the boundary entanglement entropy for a boundary subregion $R$ is given by the bulk entanglement between two regions - the region inside the corresponding RT surface in the bulk and its complement. The relevant regions are depicted in figure \ref{genfig}. This bulk entanglement entropy can be computed order by order in Newton's constant $G$, using the replica trick in the bulk, \cite{Lewkowycz:2013nqa, Faulkner:2013ana} as\footnote{For the cases where we have a $U(1)$ symmetry as for static black holes, it is easier to implement the replica trick for non-integer $n$, $n$ being the number of replicated geometries. For more general cases without any $U(1)$ symmetry, one needs to define the partition function for non-integer $n$ separately \cite{Faulkner:2013ana}.}
\begin{equation}
\label{sbulkexp}
S_{\mathrm{bulk}}(R)=S_{\mathrm{bulk},cl}(R)+S_{\mathrm{bulk},q}(R),
\end{equation}
where the first term on the right-hand side of \eqref{sbulkexp} scales as ${1}/{G}$ (or equivalently is of order $N^2$) and corresponds to the minimal area surface term\footnote{Note that the motivation of \cite{Faulkner:2013ana} was to connect the bulk entanglement entropy with the quantum correction of boundary entanglement entropy. So, these authors only studied the $S_{bulk,q}$ part, which they refer to as $S_{bulk-ent}$.}, while the second term scales as $G^0$ and corresponds to the first quantum correction to the boundary entanglement entropy.

In \cite{Faulkner:2013ana}, the quantum correction to the boundary entanglement entropy $S_{EE,q}$ is given by
\begin{equation}\label{faulkform}
S_{EE,q}=S_{\mathrm{bulk}}=-\partial_n\left(\log Z_{n,q}-n\log Z_{1,q}\right)\big{|}_{n\to 1},
\end{equation}
where $Z_{n}$ is the bulk partition function of the replicated geometry in the bulk.

Taking these results into account, we now proceed to state our observations. In the path-integral language, the decomposition of \eqref{sbulkexp} can be realized by writing the full bulk partition function $Z^{\mathrm{bulk}}$ as
\[
Z^{\mathrm{bulk}} = W^{\mathrm{bulk}} + W^{\mathrm{bulk}}_{\mathrm{eff}},
\]
where $W^{\mathrm{bulk}}$ denotes the classical action. This gives the classical part of bulk entanglement entropy. It is essentially the same minimal area surface term that appears in Wald's treatment of the first law \cite{Wald:1993nt}. $W^{\mathrm{bulk}}_{\mathrm{eff}}$ is the one-loop effective bulk action which gives $S_{\mathrm{bulk,q}}(R)$.

In the framework of replicated $n$-fold geometries $\hat{g}_n$, the full density matrix $\hat{\rho}'_n$ is given in terms of a bulk time dependent Hamiltonian $H_{\tau,\mathrm{full}}$ which generates the time translation along the Euclideanized time $\tau$ direction \cite{Lewkowycz:2013nqa}. That is,
\begin{align}
\hat{\rho}'_n&=e^{-\int_{0}^{2\pi n}H_{b,n,\mathrm{full}}}\nonumber\\  &=e^{-\int_{0}^{2\pi n}\left(H_{b,n,cl}+H_{b,n,q}\right)}=\hat{\rho}'_{n,cl}\cdot \hat{\rho}'_{n,q} \, ,\label{denmatexpn}
\end{align}
where the subscripts $b$, $n$, $cl$ or $q$ above respectively suggest that the associated Hamiltonian $H$ is in the bulk, in $n$-deformed spacetime and it is either classical or quantum  (order by order in the $G$ expansion). These classical and quantum parts give rise to the classical and quantum parts of the corresponding bulk entanglement $S_{\mathrm{bulk}}$. From the above, it is easy to see for diagonal density matrices that by inserting  the expression \eqref{denmatexpn} into the von Neumann bulk entanglement entropy,  $S_{\mathrm{bulk}}$ divides into  classical and quantum parts as in \eqref{sbulkexp}, i.e.
\begin{align}\label{bulkee}
S_{\mathrm{bulk}}=&-\partial_n\left[\log\mbox{Tr}(\hat{\rho}'_{n,cl})-n\log\mbox{Tr}(\hat{\rho}'_{1,cl})\right]\nonumber\\
&-\partial_n\left[\log\mbox{Tr}(\hat{\rho}'_{n,q})-n\log\mbox{Tr}(\hat{\rho}'_{1,q})\right]+\dots\nonumber\\
=&S_{\mathrm{bulk},cl}(R)+S_{\mathrm{bulk},q}(R)+\dots \, ,
\end{align}
where the dots denote  terms that are local integrals on the RT surface. When only the background metric has a non-zero vacuum expectation value, there are two other terms that in principle can contribute to the $O(G^0)$ correction to the boundary entanglement entropy. These come from a change in area due to the back-reaction on the classical background and from general higher derivative terms, respectively.\footnote{In addition,  counterterms may be necessary to ensure finite entanglement.
} For our present purpose we do not consider the higher derivative terms in the bulk action. 

Our key observation in this section will be the term-by-term matching of the expansions \eqref{sbulkexp} or \eqref{bulkee} to an analogous expansion of the Fisher information metric, namely, 
\begin{equation}\label{GRexp}
G_{F,\lambda\lambda}=G_{F,cl}+G_{F,q}.
\end{equation}
Once again, the subscripts $cl$ and $q$ denote the classical and quantum parts.

To begin with, let us consider a perturbation of the background metric $g^{(0)}$ of the form
\be
\label{met-pert}
g=g^{(0)}+\delta g^{(0)} + h.
\ee
Here we consider two different kinds of perturbations of the bulk metric. $h$ is an $O(\sqrt{G})$ quantum fluctuation, while $\delta g^{(0)}$ takes into account the $\lambda$ variation.
Expanding the right-hand side of \eqref{einstein} in powers of $\lambda$ according to \eqref{met-pert}, and inserting the expansion back into \eqref{result-Lashkari}, we obtain
\begin{align}
\frac{\partial^2}{\partial\lambda^2}S^{\mathrm{bdy}}_{\mathrm{rel}}(\rho_\lambda||\rho_0)\Big{|}_{\lambda=0} =& \, {\cal E} - \int_{\Sigma}\xi^\mu T^{\mathrm{grav},(2)}_{\mu\nu}(\delta g^{(0)}) v^\nu  \nonumber \\ 
&- \left[\int_{\Sigma}\xi^\mu T^{\mathrm{grav},(2)}_{\mu\nu}(h) v^\nu  + \int_{\Sigma}\xi^\mu T^{\mathrm{matter},(2)}_{\mu\nu}(g) v^\nu\right] \nonumber \\ 
&+ \mbox{boundary terms} \, .
\label{class-quant}
\end{align}
This gives a clean separation of classical and quantum contributions in the Fisher metric and also the classical and quantum contributions in the leading order canonical energy in the perturbed background. The first two terms on the right-hand side of  \eqref{class-quant} are classical ($\mathcal{O}(1/G)$) contributions, with the second term arising from \eqref{can-en} upon the second order variation in $\delta g^{(0)}$. These two classical terms are the same as the right hand side of \eqref{result-Lashkari}. The superscript $(2)$ signifies the fact that all the variations are of second order in $\delta g^{(0)}$ and $h$.\footnote{Note that the first order variation in either case vanishes by virtue of linearized equation of motion.} The remaining terms in the bracket are quantum corrections.\footnote{Also note that the matter part of the stress tensor only appears at the quantum level as classically for empty AdS, the contribution is identically zero.} Furthermore, it was shown that the boundary terms can be taken care of through a suitable choice of gauge as pointed out in \cite{Hollands:2005wt}.

Now combining \eqref{fisher-rel1} and \eqref{class-quant} enables us to schematically write
\begin{align}
\label{G-sep-C-sep}
\frac{\partial^2}{\partial\lambda^2}S_{\mathrm{rel}}^{\mathrm{bdy}}(\rho_\lambda||\rho_0)\Big{|}_{\lambda=0} &= G_{F,\lambda\lambda} \nonumber \\
&= G_{F,cl}+G_{F,q},
\end{align}

Thus from \eqref{fisher-rel1} and \eqref{class-quant}, we obtain an expansion of Fisher information metric at order by order in Newton's constant and their respective connections with the classical and quantum part of the canonical energy. 
Bearing in mind \eqref{fid-rel1}, for the special case of commuting density matrices, this also calls for a decomposition analogous to  \eqref{G-sep-C-sep}  for the {\it reduced} fidelity susceptibility, as
\be
\label{GR-sep-C-sep}
\frac{\partial^2}{\partial\lambda^2}S_{\mathrm{rel}}^{\mathrm{bdy}}(\rho_\lambda||\rho_0)\Big{|}_{\lambda=0} = G_{R,cl}+G_{R,q} \, .
\ee

In the next subsection, we further develop this connection, where we relate $G_{F,q}$ (and $G_{R,q}$ for the restricted class of states corresponding to commuting density matrices) to the bulk modular Hamiltonian and hence to bulk entanglement entropy. 

\subsection{Canonical energy and bulk modular Hamiltonian}\label{sec:rel-ent}

Here we complete our arguments by invoking the fact that the quantum correction to canonical energy (the bracketed term in the second and third lines of \eqref{class-quant}) is essentially the same as the bulk modular Hamiltonian ${\mathcal{H}_R}^{\mathrm{bulk}}$ that appears as the first quantum correction to the boundary modular Hamiltonian \cite{Jafferis:2014lza}, \cite{Jafferis:2015del}
\begin{equation}\label{modhamexpjaff}
{\mathcal H_R}=\frac{\mbox{Area}(\gamma)}{4G}+{\mathcal H_R}^{\mathrm{bulk}}+\dots \, .
\end{equation}

This is the operator equivalent\footnote{This is possible by noting the connection between the entanglement entropy and the modular Hamiltonian via the density matrix as in \eqref{mod-Ham}.} of the expansion of boundary entanglement entropy at order by order in $G$, namely
\begin{equation}\label{SEEexpJaff} 
S_{EE}=\frac{\mbox{Area}(\gamma)}{4G}+S_{EE,q}+\dots \, .
\end{equation}
Thus the results \eqref{class-quant},  \eqref{G-sep-C-sep}, \eqref{GR-sep-C-sep}, \eqref{modhamexpjaff}, \eqref{SEEexpJaff} clearly suggest that the quantum Fisher information metric and equivalently the reduced fidelity susceptibility for the mentioned restricted class of states can indeed be understood as sum of two terms as in \eqref{GRexp}, namely a leading semiclassical term and a subleading quantum term. In this division we are simply keeping track of the orders $G^{-1}$ and $G^{0}$, respectively. 

For example, if we just focus on the quantum part, i.e. the $\mathcal{O}(G^{0})$ part, we see from \eqref{G-sep-C-sep} and \eqref{class-quant} that 
\begin{eqnarray}
\label{propBeq}
G_{F,q} &=& -\left[\int_{\Sigma}\xi^\mu T^{\mathrm{grav},(2)}_{\mu\nu}(h) v^\nu+\int_{\Sigma}\xi^\mu T^{\mathrm{matter},(2)}_{\mu\nu}(g) v^\nu\right] \nonumber \\
&=& S_{EE,q} \, ,
\end{eqnarray}
where the last quantity arises from the quantum canonical energy and is equal to the modular Hamiltonian in the bulk $\mathcal{H}_{R}^{\mathrm{bulk}}$ \cite{Jafferis:2014lza}, \cite{Jafferis:2015del}. 

Thus following the separation in the classical and the quantum parts in \eqref{class-quant}, we conclude that while the classical part of the Fisher information metric $G_{F,\lambda\lambda}$ is given by the classical part of canonical energy in agreement with \cite{Lashkari:2015hha}, the quantum part of it can be thought of as a dual to the bulk modular Hamiltonian. The same conclusion holds for the reduced fidelity susceptibility, however only for the restricted class of states leading to commuting density matrices.

Finally, for the excited states discussed above in section \ref{subsec:scalerpert} due to marginal scalar perturbations, we point out that our second proposal also goes through. This can be understood by noting the results of \cite{Beach:2016ocq}, who proved that for such perturbations, the Fisher information becomes canonical energy in the bulk. Of course the connection between Fisher information and the canonical energy is what enables us to provide a further proof of the second part of our proposal.

\section{Conclusions and outlook}\label{sec:conclusion}

In the first part of this work we have proposed a holographic dual of Fisher information metric for  mixed states. In all the cases that we consider, this is always given by a regularized (i.e.~finite) volume contained under the RT surface in the bulk. This also serves as a holographic dual for the reduced fidelity susceptibility but for a restricted class of states, namely, when the density matrices commute before and after perturbation, i.e when the states are effectively classical. At least for this class of states we can compare our result for a previous proposal for the holographic dual of the reduced fidelity susceptibility given in \cite{Alishahiha:2015rta} in terms of holographic complexity, namely, the leading term in the volume under the RT surface. However, the proposal given there suffers from the following shortcomings. As we mentioned before,  for classical (or effectively classical) states, fidelity susceptibility is physically the same as Fisher information, which is defined by the second order variation of relative entropy. Now relative entropy for a mixed state is always UV-finite. Hence it is hard to justify that holographic complexity, which is UV-divergent, should be its bulk dual.  UV-convergent behaviour was also advocated from a purely field theory computation in \cite{Lashkari:2014yva}, at least for free theories and conformal field theories with large central charge. Furthermore, the second-order variation of relative entropy was computed explicitly \cite{Lashkari:2015hha, Blanco:2013joa}, and its behaviour differs significantly from that of holographic complexity as proposed in \cite{Alishahiha:2015rta}. On the other hand, as we have shown, our proposal for the holographic dual of reduced fidelity susceptibility for those states meets both requirements by construction. Our proposal for the bulk dual is  similar in spirit to the recently proposed idea of complexity of formation \cite{Chapman:2016hwi, Carmi:2016wjl, Reynolds:2016rvl}, which measures the relative complexity between two states. One natural question might then be whether this is also related to the computational complexity as discussed in \cite{Brown:2015bva, Brown:2015lvg, Susskind:2014moa}. In fact, its connection with the relative entropy is reminiscent of the definitions of complexities used in \cite{Nielsen05,Gomez99}.

One naive intuition to justify the relation to  computational complexity comes from the positivity of both reduced fidelity susceptibility and  Fisher information. As already mentioned in \cite{Lashkari:2016idm}, the identification of Fisher information with the Hollands-Wald canonical energy  \cite{Lashkari:2015hha,Hollands:2012sf} implies the positive energy theorem for asymptotically AdS spacetime. Our result hints at an alternative way to view the derivation of the positive energy theorem for asymptotically AdS spacetimes in terms of positivity of reduced fidelity susceptibility for  mixed states. A suggestion is  to interpret this positivity of canonical energy (in our case, we consider the canonical energy associated to the bulk Rindler wedge corresponding to the spherical boundary region $R$) as the transition from a reference vacuum state to a more complex excited state. In other words, the monotonically increasing nature of reduced fidelity susceptibility mimics that of  computational complexity. However, a full justification behind such a connection is yet to be understood. Work in this direction is in progress and we hope to report on this generalization in near future. In particular, we are working on a computation within quantum field theory to reproduce the scaling with $R^{2d}$ as in \eqref{holFisher1} and \eqref{holFisherd} for stress-tensor perturbations. Even within our computation and proposal, we pointed out various subtleties relating to the difference in coefficients in the volume-Fisher relation, and it will be interesting to investigate whether we can say more about them in a better unified manner and concretize our porposal.

The second part of our proposal relates the quantum contribution to Fisher information or reduced fidelity susceptibility to bulk entanglement. The latter has been argued to be instrumental in understanding the reconstruction of the bulk points inside an entanglement wedge in terms of local operators in the boundary CFT through modular evolution \cite{Jafferis:2015del}. We expect our proposed duality might add an useful component towards a concrete study in this direction.

There are many other important issues that need to be investigated in future for a complete understanding of the proposed connections. We already mentioned  the quantum information origin of (mixed state) holographic complexity itself, which is missing so far in the literature. It will also be interesting to understand how our construction changes for more complicated boundary states, such as subregions with arbitrary shapes or thermofield double geometries. A generalization to multi-dimensional parameter space and a covariant generalization of our proposal also deserve a closer look.

\bigskip
\goodbreak
\centerline{\bf Acknowledgements}
\noindent
We would like to thank Raimond Abt, Dean Carmi, Dan Kabat, Nima Lashkari, Juan Maldacena, Christian Northe and Ignacio Reyes for helpful discussions. We particularly thank Mohsen Alishahiha and Tadashi Takayanagi for numerous helpful discussions and comments on the manuscript. We would also like to thank Ayan Mukhopadhyay for comments on an earlier version of the manuscript. Last, but not the least, we would like to thank Nina Miekley for collaborating on the results presented in section \ref{subsec:scalerpert} and several useful discussions. The work of SB is supported by the Knut and Alice Wallenberg Foundation under grant 113410212. The work of DS is supported by the ERC grant `Self-completion'.
\appendix

\providecommand{\href}[2]{#2}\begingroup\raggedright


\endgroup

\end{document}